\documentclass[prb,twocolumn,aps,showpacs]{revtex4}
\usepackage{graphicx}

\begin{document}
\title{Comment on ``Frustrating interactions and broadened magnetic
interactions in the edge--sharing CuO$_2$ chains in
La$_5$Ca$_9$Cu$_{24}$O$_{41}$''}
\author{R.~Leidl}
\author{W.~Selke}
\affiliation{Institut f\"ur Theoretische Physik, Technische Hochschule,
52056 Aachen, Germany}
 
\begin{abstract}
  Using Monte Carlo techniques, we show that the
  two--dimensional anisotropic Heisenberg model reproducing
  nicely inelastic neutron scattering measurements on 
  La$_5$Ca$_9$Cu$_{24}$O$_{41}$ (Matsuda
  \textit{et al.} [Phys.\ Rev.\ B {\bf 68}, 060406(R) (2003)]) seems
  to be insufficient to describe correctly measurements
  on thermodynamic quantities like the magnetization or the susceptibility.
  Possible reasons for the discrepancy are suggested.   
\end{abstract}


\maketitle
 
Recently, Matsuda \textit{et al.}\cite{mat} reported results
of ca\-reful inelastic neutron scattering experiments on
La$_5$\linebreak[0]Ca$_9$Cu$_{24}$O$_{41}$. From standard
spinwave analysis, it is concluded that the experimental
data can be well reproduced by a two--dimensional anisotropic
spin-1/2 Heisenberg model with antiferromagnetic
nearest--neighbor interactions, $J_{c1}$ (= 0.2meV), and ferromagnetic
next--nearest--neighbor couplings, $J_{c2}$ ($= -0.18$meV), between the
Cu ions in the CuO$_2$ chains, as well as antiferromagnetic
couplings $J_{ac1}$ (= 0.681meV) and $J_{ac2}$ ($=0.5J_{ac1}=0.3405$meV)
between nearest and next--nearest neighboring Cu ions
in adjacent chains, respectively (see Fig.~1 in Ref.~\onlinecite{mat}).
The uniaxial anisotropy of the spins along the $b$ axis is written in the
form of a single--ion interaction
$D$ (= $-0.211$meV), summing over the anisotropy contributions
of the different couplings.

This model reproduces the measured dispersion relations
along the CuO$_2$ chains, i.e. along the $c$ axis, and
perpendicular to these chains, along the $a$ axis. It
is supposed to describe, at
low temperatures, the long--range ferromagnetic order in
the chains and an antiferromagnetic order perpendicular to
the chains.  

We performed Monte Carlo simulations on the
model in its classical limit with unit vectors
as the spins, where the uniaxial anisotropy along
the $b$ axis is
described either by a single--ion
anisotropy,\cite{mat} $D \sum (S_j^z)^2$, or
by anisotropic exchange interactions distributing the total anisotropy in
various ways among the different couplings. We used exactly
the interaction strengths determined 
in Ref.~\onlinecite{mat}. To
compare the simulational data to previous measurements \cite{amm}
on the magnetization and specific heat
of La$_5$Ca$_9$Cu$_{24}$O$_{41}$, we 
included external fields $H_{b,c}$ along the $b$ and $c$
axes, and we recorded especially sublattice
magnetizations, the total magnetization, the susceptibility, and the
specific heat. We took care that reliable
equilibrium data were obtained, using at least $5\times10^5$ Monte
Carlo steps per spin in each run. We studied finite--size
effects, simulating square lattices with the linear dimension ranging
from 10 to 100. The square lattice is constructed
in such a way that its principal axes 
are given by the two directions along which $J_{ac1}$ acts 
between a reference site and the two sites in one of the
adjacent chains. Thence the relatively strong coupling
$J_{ac1}$ connects nearest--neighbor spins on that lattice,
and the CuO$_2$ chains run diagonally through the lattice.
This geometry shows most directly the relation of the model
by Matsuda \textit{et al.} to much studied anisotropic Heisenberg models
with nearest--neighbor interactions on square lattices.

Our main findings may be summarized as follows:

(i) For vanishing field and at low temperatures, one
obtains, indeed, antiferromagnetic order, i.e.\ the
model describes a ferromagnetic ordering of the spins
along the CuO$_2$ chains and an antiferromagnetic ordering
perpendicular to the chains, due to the fairly strong 
antiferromagnetic couplings $J_{ac1}$ and $J_{ac2}$. The
frustrated intrachain couplings $J_{c1}$ and $J_{c2}$, tending
to compensate each other, play only a minor role, connecting
next--nearest and more distant spins on the square lattice. The
transition from the antiferromagnetic to the
para\-magnetic phase occurs at about $T_N= 11.5 \pm 0.5$K (shifting
to somewhat lower temperatures when assigning the uniaxial
anisotropy mainly to the intrachain couplings). This apparent
agreement with the experimentally observed transition
temperature of $10.5$K (Ref.~\onlinecite{mat})
resp.\ $9.7$K (Ref.~\onlinecite{amm})
is, however, misleading. Quantum fluctuations are known
to reduce the transition temperature in closely related
two--dimensional anisotropic spin-1/2 antiferromagnetic quantum Heisenberg
models by a factor of about 1.5 in the extreme Ising limit, as
compared to $T_N$ in the corresponding classical 
models with unit vectors; for weaker anisotropy the factor tends
to increase.\cite{cuc,ser} Therefore we conclude that the
transition temperature of the anisotropic Heisenberg model \cite{mat} is
expected to be too low compared to the measured $T_N$.

\begin{figure}
  \includegraphics[width=8.6cm]{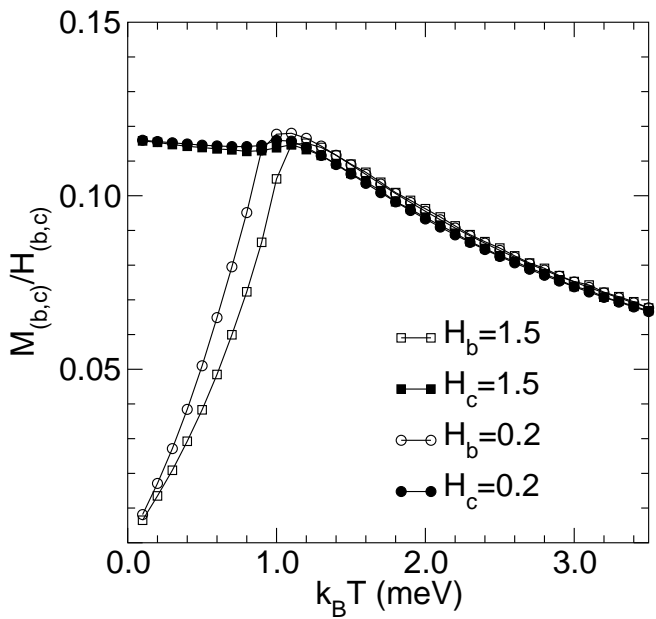}
  \caption{Total magnetization $M_{(b,c)}$ divided
  by the strength of the differently oriented fields $H_{(b,c)}$ as
  a function of temperature as obtained from Monte Carlo
  simulations of the classical version of the Hamiltonian
  by Matsuda \textit{et al.} with a single ion anisotropy,
  Ref.~\onlinecite{mat},
  for a system with 60 $\times$ 60 spins. Note
  that $H_{(b,c)}$ is expressed in units of meV.}
  \label{fig1}
\end{figure}

(ii) Well above the transition temperature, say, at 
$1.5\,T_N < T < 3.0\,T_N$, the ratio $r$ of the static
susceptibilities (defined by the
magnetizations divided by the magnetic fields) along
the $b$ and $c$ axes, resp., is found to be almost independent
of temperature, both in experiments \cite{amm}
and in the simulations, see Fig.~\ref{fig1}.
For the anisotropic Heisenberg model, we find that the ratio is
also largely independent of the strength of the field (considering fields
smaller than the spin--flop field at zero
temperature, $H_b^{\text{sf}}$= 1.81meV), with
$r \approx 1.02$ (approaching one at higher temperatures). The
measured ratio is appreciably larger,  $r \approx 1.3$, see Fig.~1 in
Ref.~\onlinecite{amm}. This discrepancy may
be, however, misleading because the ratio $r$ obtained
in the simulations has to be multiplied by the ratio of the
squares of the corresponding $g$--factors, $(g_b/g_c)^2$, for
a correct comparison with
the experiments.\cite{klin,mat2} Indeed, the $g$--factor is
anisotropic, with $g_b/g_c \approx 1.14$ for this
material.\cite{klin,exp} Thence the simulational
findings for the ratio $r$ of the magnetizations in differently
oriented fields on the anisotropic
Heisenberg model with a rather weak anisotropy seems to be
consistent with the pertinent experiments.  

\begin{figure}
  \includegraphics[width=8.6cm]{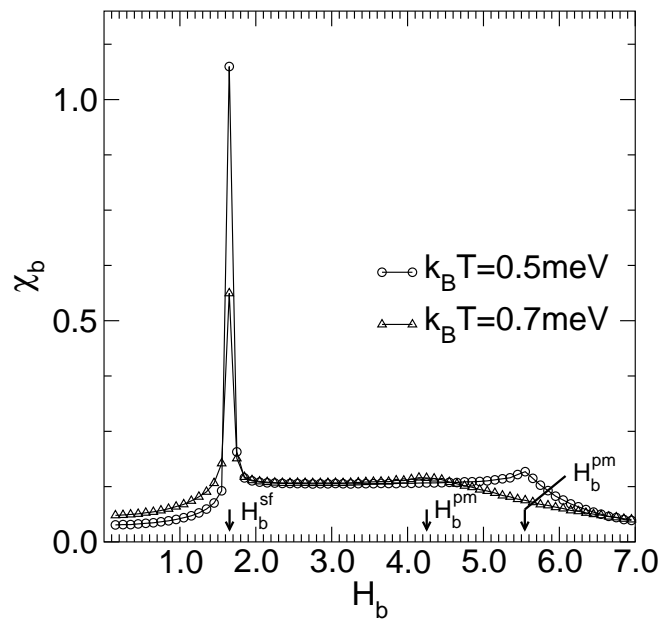}
  \caption{Susceptibility $\chi_b=\partial M_b/\partial H_b$
  at $k_BT=$ 0.5meV and 0.7meV as determined in
  simulations of the classical version of the Hamiltonian
  by Matsuda \textit{et al.} for a system with 60 $\times$ 60 spins.}
  \label{fig2}
\end{figure}

(iii) Applying the field along the $b$ axis, we evaluated the
susceptibility $\chi_b=\partial M_b/\partial H_b$
at temperatures below $T_N$. Typically, at constant temperature,
$\chi_b$ displays as a function of $H_b$ a strong, delta--like peak
at $H_b^{\text{sf}}$, signalling the transition from
the antiferromagnetic to the spin--flop phase and, at a larger
field, $H_b^{\text{pm}}$, an additional weak maximum indicating the transition
from the spin--flop phase to the
paramagnetic phase, see Fig.~\ref{fig2}.
The spin--flop field $H_b^{\text{sf}}$ is nearly independent
of temperature below about $0.7\,T_N$, decreasing then rather sharply to zero
as $T$ approaches $T_N$. The
upper transition field, $H_b^{\text{pm}}$, decreases quite strongly
with increasing temperature already at low temperatures.  These
findings on the anisotropic Heisenberg model are in marked contrast
with the observed behavior \cite{amm,klin} of
$\chi_b$ for La$_5$Ca$_9$Cu$_{24}$O$_{41}$. Experimentally, one 
also obtains two anomalies in $\chi_b$, but a sharp, cusp--like
singularity with a rather low maximal
height at a small field followed by
a broad and higher maximum at larger fields. The location of the upper
characteristic field depends, below
$T_N$, only weakly on temperature. At the lower characteristic
field the breakdown of antiferromagnetic order is observed in
neutron scattering experiments;\cite{klin} this field goes to
zero as one approaches $T_N$. Experimentally, there seems to be
no evidence for a transition from the antiferromagnetic phase
to a spin--flop phase at the lower
characteristic field. The origin and nature of that transition has
still to be clarified. At any rate, the results
on $\chi_b$ for the anisotropic Heisenberg model 
deviate clearly from the measured behavior.

From the comparison between the thermodynamic
measurements \cite{amm,klin} on La$_5$Ca$_9$Cu$_{24}$O$_{41}$
and the simulational data on the classical
version of the model suggested in Ref.~\onlinecite{mat}, it
follows that the model, describing nicely the dispersion relations, seems
to be incomplete, missing important
ingredients of the real material. Perhaps, as
already indicated before,\cite{mat} the holes induced by
the Ca ions may play an important role. A first, simplified
modelling of that aspect in the framework
of an Ising model has been proposed
recently. \cite{sel,hol} Furthermore, the
couplings of spins in neighboring $ac$--planes may
be relevant. Of course, extending the model
of Matsuda \textit{et al.}\cite{mat}
to three dimensions is not expected to resolve the discrepancy
about the presence of the spin--flop phase. Finally, structural
distortions in La$_5$Ca$_9$Cu$_{24}$O$_{41}$ may even call for a
description going beyond pure spin models.

\begin{acknowledgments}
  We thank B.\ B\"uchner and  R.\ Klingeler for very useful
  discussions as well as M.\ Krech and M.\ Matsuda for
  helpful information.
\end{acknowledgments}

\end{document}